# A Review Of Mass Estimates For Galactic Dark Matter Objects

**Abstract:**   Empirical mass estimates for galactic dark matter objects, published between December 1991 and May 1999, are presented in tabular and graphical forms.  Trends in the data are identified and uncertainties are discussed.  Similarities among various stellar and dark matter mass functions are noted, and a possible identification of the galactic dark matter objects is suggested.

Robert L. Oldershaw

Amherst College

Amherst, MA 01002

USA

rlolders@unix.amherst.edu

## 1. Introduction

In a sequel to a review of empirical mass estimates for galactic dark matter objects (GDMOs) which covered the period from December 1991 to December 1996 (Oldershaw, 1998), more recent mass estimates from the period December 1996 to May 1999 are reviewed. Uncertainties, trends and issues related to the identification of the dark matter (DM) are discussed. In Section 2 the mass estimates are tabulated, and graphed as a function of publication date. Various sources of uncertainty in determining the masses of GDMOs are identified in Section 3, while trends in the data are explored in Section 4. A possible resolution to the enigmatic identity of the GDMOs is suggested in Section 5.

## 2. The Data

Table 1 lists mass estimates for dark matter objects, their date of publication, comments and references. The mass estimates have been collected from published papers and preprints. The author has attempted to provide a complete data set for the December 1996 to May 1999 period, but in such a rapidly developing field, that is not a simple matter. What can be said is that the data set reported in Table 1 is nearly complete and highly representative of the results presented during the time period under review. There is an inevitable amount of heterogeneity in the data: some estimates are based on hundreds of events while others are based on less than ten events. Some authors reported mass ranges while others cited average, median, or most typical values. However, since we are interested mostly in general trends and approximate mass estimates (given the substantial uncertainties in the estimates), this is not a serious problem. Previously reviewed data from the December 1991 to December 1996 are included for comparison purposes.

Figure 1 displays the full mass range for currently viable DM candidates along the x-axis in units of log $M_o$. The y-axis depicts the number of months since November 1991 when the first microlensing results were published. The three columns on the left hand side of the graph represent estimated mass ranges for the currently favored particle DM candidates: $10^{-6}$ ev to $10^{-4}$ ev for axions, 2 ev to 30 ev for massive neutrinos, and 10 Gev to 500 Gev for neutralinos (Dodelson *et al*, 1996). The two vertical lines at 0.145 $M_O$ and $7 \times 10^{-5}$ $M_O$ represent predicted GDMO mass values derived from a fractal cosmological model possessing discrete self-similarity (Oldershaw, 1987), which will be discussed in Section 5. Figure 2 is an expanded version of the right hand portion of Figure 1. Dotted lines in both figures show the December 1996 division between previously reviewed data and more recent mass estimates.

**3. Uncertainties In GDMO Mass Estimates**

At present there are still unavoidable uncertainties involved in deriving mass estimates for GDMOs, and these should be mentioned here.

**3.1. Degeneracy**

Microlensing observations record the Einstein crossing time ($t_E$) for an event, and this is correlated with the mass of the lens (m), the distances between components of the lens system ($d_i$), and the transverse velocity ($v_t$) of the lens (Paczynski, 1986). Unfortunately, the degeneracy in m, $d_i$ and $v_t$ preclude a straightforward measurement of m. Instead, one calculates a most probable value (or range) of m, given the most likely values (or ranges) of $d_i$ and $v_t$. A reanalysis of the OGLE group results demonstrates that factor of 2 to 3 uncertainties in mass estimates can be attributed to the degeneracy problem (Alard, 1999). There is some hope that future experiments permitting parallax measurements will break the degeneracy and give us increasingly accurate $m_{DM}$ values (Gould, 1999).

**3.2. Blending**

In an ideal micolensing event only one source object is lensed, but actual events may involve binary sources or a projected superposition of sources at different distances. This blending of more than one source per event must be taken into account, and introduces statistical uncertainties (Wozniak and Paczynski, 1997; Sutherland, 1999).

**3.3. Galactic Model Uncertainties**

There are several alternatives to the standard microlensing model wherein DM lenses form a spherical halo around the Galaxy and stellar sources reside in the Magellanic Cloud galaxies (Sadoulet, 1999; Sutherland, 1999). Gyuk and Gates (1999) have discussed the possibility that a "thick disk" model might be superior to the spherical halo model. The estimated average GDMO mass ($<m_{DM}>$) is about 0.4 $M_o$ in the latter case, but $<m_{DM}>$ is lowered to ≈ 0.27 $M_o$ in the thick disk model. Another analysis (Gyuk and Gates, 1998) suggests that going from a standard halo model to a rotating thick dick model can decrease $<m_{DM}>$ to a limit of about 0.15 $M_o$ to 0.20 $M_o$. Evans et al (1998) have proposed that the lensing objects might be located in a thinner Galactic disk which is strongly warped in its outer parts, thus bringing disk lenses into our line-of-sight to the Magellanic Clouds. Two other possibilities are that the lenses reside in a hitherto unseen dwarf galaxy that lies between us and the Large Magellanic Cloud (Zhao, 1996), or that tidal debris from the LMC might account for the halo lensing results (Zhao, 1998). It is also possible that some of the halo events are actually due to "self-lensing", wherein *both* the lenses and sources are located within the Magellanic Cloud galaxies (Sahu, 1994). There is evidence that this indeed might be the case for several well-observed events (Bennett, et al, 1996; Alcock, et al, 1997a; Sahu and Sahu, 1999). In the case of bulge events, it has been suggested (Zhao, et al, 1995) that a bar-shaped Galactic bulge might explain the anomalously high optical density derived from microlensing observations in that direction. However, the viability of this model has been challenged (Sevenster, et al, 1999). Mass estimate uncertainties for planetary-mass GDMOs are still very large (Pelt, et al, 1998; Rhie et al, 1999).

It is sound scientific practice to explore alternative explanations for the results of microlensing observations, but it would seem that the model requiring the fewest *ad hoc* assumptions, in this case the standard model, should be preferred for the present. Larger numbers of events and future experiments using modified strategies (Gould, 1992, 1999; Di Stefano and Scalzo, 1999; Zhao, 1999) will help to determine more accurately the masses, abundances and distributions of the GDMOs.

**4. Trends in the Data**

**4.1. Previously identified patterns**

The review of GDMO mass estimates reported between November 1991 to December 1996 identified three main patterns in the data (Oldershaw, 1998).

(1) Stellar-mass GDMOs with most probable masses in the 0.1 $M_o$ to 0.5 $M_o$ range appeared to contribute a significant fraction (10 to 100%) of the galactic halo DM.

(2) A second, and apparently distinct, mass peak of planetary-mass GDMOs was indicated in the less well-defined range of $10^{-7}$ $M_o$ to $10^{-3}$ $M_o$.

(3) There was an absence of evidence for objects below the planetary-mass lower limit ($\approx 10^{-7}$ $M_o$). Surprisingly, there was no evidence for particle DM candidates, in spite of extraordinary efforts to detect them.

**4.2. Trends in the more recent data**

The newer data from December 1996 to May 1999 are, for the most part, consistent with the older data, and strengthen the case for a combination of stellar-mass and planetary-mass DM populations. After seven years of microlensing experiments the galactic dark matter is still one of the most important puzzles in astrophysics. On the other hand, much has been learned during this period, and the most likely, although still tentative, conclusions that can be drawn from the existing data, as discussed in earlier sections, are listed below.

### 4.2.1. Stellar-Mass GDMOs

As can be readily seen in Figs. 1 and 2, the more recent data provide further confirmation of a substantial population of stellar-mass GDMOs with estimated masses in the 0.1 $M_o$ to 0.5 $M_o$ range. On the basis of over 200 bulge events, Zhao and de Zeeuw (1998) find that the "current best estimates of the lens masses, which use realistic models for the Galactic bulge or bar, indicate masses near 0.15 $M_o$ ...". Evans and Gyuk (1998) report "an average mass of around 0.3 $M_o$" for halo lenses. Jetzer (1998) concludes that "the value of ~ 0.1 $M_o$ looks as the most realistic estimate to date". On the basis of 14 halo events, the MACHO group cites a "most probable MACHO mass of 0 .5 (+0.3, -0.2) $M_o$" (Sutherland, 1999). A rough average of all reported mass estimates for stellar-mass GDMOs is approximately 0.2 $M_o$. A small, but significant, number of microlensing events are associated with masses > 0.5 $M_o$. Mass estimates in the 0.01 $M_o$ to 0.1 $M_o$ range are quite rare. The MACHO group finds no event durations less than 20 days for the halo (Sutherland, 1999).

Sandoulet (1999) comments: "If we assume that the MACHOs are distributed in the same way as the galactic halo, they may represent a fraction of the halo density between 10 and 100% (Fig. 1). Although the compatibility with 100% may superficially indicate that the dark matter problem is solved, this interpretation encounters *the serious difficulty* that the mass of individual lenses would be typically *one third of a solar mass*. These objects are not brown dwarfs. They cannot be ordinary stars ... [or] white dwarfs [which would require] an artificial initial mass function, an uncomfortable age of more than 18 billion years, and a totally unknown formation mechanism. We are led to question the assumed distance and velocity distributions. Four types of models have been proposed ..." (emphasis added). Unfortunately, this argument does not mention the possibility that the GDMOs are a previously unknown class of objects such as primordial black holes with masses on the order of 0.3 $M_o$. The dark matter problem has been an enigma for decades and we need to be broad-minded about its possible solutions. Science requires sufficient empirical justification before rejecting "face-value" answers. Such a requirement is especially true in this case since one cosmological model predicted that the major GDMO mass peak would be found at ≈ 0.2 $M_o$ and would be comprised by ultracompact objects (Oldershaw, 1987), as will be discussed in Section 5.

### 4.2.2  Planetary-Mass GDMOs

There is some further empirical support for a separate planetary-mass GDMO population within the mass range of $10^{-6}$ $M_o$ to $10^{-2}$ $M_o$. On the basis of quasar light curves, Hawkins (1998, 1999) has argued that most quasars show signs of microlensing by a population of compact planetary-mass objects, with primordial black holes cited as being the most likely candidates. In what may be a harbinger of things to come in the area of microlensing by planetary-mass lenses within our galaxy, Rhie et al (1999) have recently reported a very high amplification MACHO event with an associated mass range of ~$10^{-5}$ $M_o$ to ~$10^{-4}$ $M_o$. Pelt et al (1998) have found *possible* evidence for planetary-mass microlensing ("down to about $10^{-5}$ $M_o$") in quasar 0957+561. The *possibility* of planetary-mass microlensing has also been found in the BL Lac 0235+164 (Kraus et al, 1999) and the blazar S5 0716+71 (Sagan et al, 1999). The MACHO project has identified 22 microlensing events, out of a total of about 300 events, that appear to involve planetary-mass companions (Becker et al, 1998). Statistical analyses suggest a large population of such objects. Moreover, there has been an increasing number (25 at last count) of extra-solar planets discovered in recent years using more conventional techniques (Lissauer, 1999). This includes two planets with masses on the order of $10^{-5}$ $M_o$ orbiting the 6.2 millisecond pulsar PSR B1257+12 (Wolszczan and Frail, 1992; Wolszczan, 1994). Several groups are conducting experiments designed to observe planetary-mass GDMOs, including the MPS, MOA, EROS, and PLANET collaborations (Rhie et al, 1999). Walker (1999) has proposed methods for determining not only the masses, but also the *compactness*, of planetary-mass lenses. A definitive empirical verdict on the existence and extent of this DM population should emerge within 5-10 years.

### 4.2.3. Particle DM?

One potential novelty in the more recent DM mass estimates is the appearance of tentative evidence for two particle DM candidates. Firstly, the Super-Kamiokande collaboration has reported (Fukuda, *et al*, 1998) "a zenith angle dependent deficit of muon neutrinos", which suggests the possibility of neutrino oscillations, which would imply that at least one type of neutrino has mass. Estimates of the most plausible mass values range from 0.07 ev to roughly 25 ev for the tau neutrino. Most physicists believe that more evidence is needed to strengthen this claim (Kestenbaum, 1998), but the estimated mass range is plotted in Fig. 1. Secondly, at the San Grasso Facility an "apparent seasonal variation in its radiation counts" has been reported, which has been interpreted tentatively as being consistent with DM particles, possibly WIMPS of about 60 proton masses (Glanz, 1999), although this claim has been greeted with considerable skepticism (Gerbier, *et al*, 1999; Glanz, 1999).

### 4.2.4. Shifting $<m_{DM}>$ Values

Another interesting trend has been the correlation between GDMO mass estimates and the designated "best-bet" candidates. The first halo microlensing event had an estimated mass of 0.12 $M_o$ (Alcock et al, 1993), and a method-of-moments analysis of the first three events suggested an $<m_{DM}>$ of about 0.144 $M_o$ (Jetzer and Masso, 1994). Based on these results and the knowledge that low-mass halo stars could not account for the events (Rieke et al, 1989; Bahcall et al, 1994), it was thought that brown dwarf stars with slightly lower masses were the "best-bet" candidates for the halo DM. As Figure 2 shows, the next three mass estimates dropped down into the 0.04 $M_o$ to 0.08 $M_o$ range, coincident with the theoretical range for the putative brown dwarfs. Subsequently, when a dearth of short duration events cast considerable doubt on the brown dwarf candidacy, there was a gradual switch to low-mass white dwarfs as the "best-bet" GDMO candidates. During this transition period, the estimated mass values increased and leveled off at about 0.5 $M_o$. Now that the white dwarf candidacy has been shown to have its own serious problems (Adams and Laughlin, 1996; Kawaler, 1996; Canal et al, 1997; Freese et al, 1999), there has been a greater dispersion in GDMO mass estimates and wider error bars, as authors recognize that GDMOs may not fit into previously known categories of astronomical objects.

### 4.2.5.  $<m_{halo}> \approx <m_{bulge}>$?

An interesting trend in the microlensing data is the *apparent* convergence of microlensing mass estimates for the bulge, halo and disk. As noted above, Zhao and de Zeeuw (1998) find a most typical lens mass of about 0.15 $M_o$ for over 200 bulge events. This value is within the error bars of the most typical value for the halo GDMOs, and three microlensing events found in the Galaxy's spiral arms have a most probable mass of about 0.3 $M_o$ (Derue, et al, 1999). In the author's previous GDMO review paper, bulge data had to be excluded because its DM content was still too hypothetical. Now, however, doubts about bulge DM have been diminished by resilient findings of bulge optical depths that exceed theoretical estimates for conventional stars by more than a factor of two (Sutherland, 1999), and by persistent differences between the observed and expected time-scale distributions of bulge events (Han and Gould, 1996; Han and Chang, 1998; Gould, 1999).

### 4.2.6.  A universal DM Mass Function?

The following list of tentative conclusions drawn from microlensing observations could apply equally well to the galactic halo or bulge lenses.

(1) There appears to be a primary, and relatively narrow, mass peak in the DMMF that is located somewhere between 0.1 $M_O$ (Jetzer, 1998) and 0.4 $M_O$ (Gould, 1999). This mass peak has a sharp decline below 0.1 $M_O$.

(2) There is some evidence for a large population of planetary-mass GDMOs. Their estimated masses range from $10^{-6}$ $M_O$ to $10^{-3}$ $M_O$, with very large uncertainties.

(3) There appears to be a significant gap in the DMMF between 0.1 $M_O$ and 0.01 $M_O$ (Sutherland, 1999), which may extend down to 0.001 $M_O$, or lower. From the standpoint of previous ideas about stellar formation, this gap was not expected and remains unexplained.

(4) There are microlensing events involving objects with masses considerably larger than those of the objects comprising the primary stellar-mass DM peak. The number of these events is relatively small, but they can cause a significant increase in $<m_{DM}>$ values. Whether they are conventional stars or GDMOs is not certain.

Therefore it has become increasingly likely that, at least to a first approximation,

$$DMMF_{halo} \sim DMMF_{bulge}. \qquad (1)$$

Occam's razor, and nearly 400 years of precedent, would lead the objective scientist to favor the hypothesis that there is one general family of DM candidates for the halo, bulge, and disk, although the abundances of different subpopulations might vary somewhat with Galactic location. The hypothesis that there are radically distinct classes of DM in different locations is logically possible, but one would want compelling empirical evidence before adopting this more complicated scenario. Tentative hints of a similar universality in stellar mass functions will be discussed below.

### 4.2.7. A universal MF?

Conventional stellar mass functions (SMFs) from a variety of environments tend to have the following typical characteristics. Many SMFs have a main peak at $\approx 0.2$ $M_O$ with a significant decline below $\approx 0.15$ $M_O$ ( D"Antona and Mazzitelli, 1994; De Marchi and Paresce, 1997, and references therein; Chabrier and Mera, 1997; Herbig, 1998; Hildenbrand et al, 1998; Pulone et al, 1999). There is a low abundance "gap" between 0.01 $M_o$ and 0.1 $M_o$, and a planetary mass peak somewhere in the $10^{-6}$ $M_o$ to $5 \times 10^{-3}$ $M_o$ range. Some have suggested that there might be a "universal" SMF (Gilmore, 1998). These characteristics are reminiscent of the halo and bulge DMMFs. Given the data available, and speaking only in terms of a *rough* first approximation, it appears that

$$\text{SMF}_{\text{typical}} \sim \text{DMMF}_{\text{halo}} \sim \text{DMMF}_{\text{bulge}}. \qquad (2)$$

If this is the case, it is an unanticipated and mysterious result, although a possible explanation will be offered below.

**4.2.8 Primordial Black Holes?**

If there are large numbers of GDMOs with masses in the 0.1 $M_o$ to 0.4 $M_o$ range, it is natural to ask what their physical state might be. Unfortunately, all of the "most reasonable" candidates within this mass range, e.g., red dwarfs, low-mass white dwarfs. brown dwarfs, low-mass neutron stars and remnant black holes, have been virtually ruled out as major constituents of the Galactic DM (Freese et al, 1999; and references therein).

Essentially by the process of elimination one is led, some would say "driven" (Freese et al, 1999), to the next most likely candidate for the stellar-mass GDMOs: primordial black holes. Hawkins (1999) argues that primordial black holes are also the most likely candidates for the planetary-mass GDMOs. A remark by Gould (1999) is relevant here: "The most viable candidates for halo lenses seem to be exotic new objects such as primordial black holes, which just happen to have the same masses as the most common stars." The last phrase notes the mysterious coincidence in stellar and dark matter mass function peaks mentioned above. At the present time there is virtually no conventional theoretical connection between primordial black hole formation and the formation of stars.

### 5. Have The GDMOs Been Identified?

Even though we have learned much from microlensing research in the last seven years, the GDMOs seem more enigmatic than ever. To put the matter simply: none of our existing conventional models or theories has anticipated, much less predicted, the strange results discussed in section 4. Moreover it is not easy to adjust these theories so that they might retrodict the results in a manner that is not uncomfortably *ad hoc*. On the other hand, one unorthodox and essentially heuristic cosmological model did predict, quantitatively and prior to the microlensing experiments, that these unexpected results would be found (Oldershaw, 1987, 1989a,b).

### 5.1. A Fractal Cosmological Model

The cosmological model under consideration is a discrete fractal model called the Self-Similar Cosmological Model (SSCM) whose basic principles, successful retrodictions, and predictions have been presented previously (Oldershaw 1987, 1989a,b; and references therein). The SSCM proposes that nature is organized into discrete hierarchical scales which exhibit self-similarity. Atomic, stellar and galactic scale systems constitute the three equally fundamental cosmological scales that are currently observable; the total number of cosmological scales is unknown at present, but is tentatively assumed to be denumerably infinite. The elemental systems of any given scale have self-similar analogues on all other scales. The heuristic scale transformation equations, which relate properties of analogues from different cosmological scales, are:

$$R_n = \Lambda R_{n-1} , \quad (3)$$
$$T_n = \Lambda T_{n-1} \text{ and} \quad (4)$$
$$M_n = \Lambda^D M_{n-1} , \quad (5)$$

where R, T and M are lengths, temporal periods and masses, respectively, of analogues on neighboring cosmological scales n and n-1. The dimensionless constants $\Lambda$ ($\approx 5.2 \times 10^{17}$) and D ($\approx 3.174$) have been determined empirically, and subsequently tested against a diverse selection of retrodictive challenges (Oldershaw, 1989a). The discrete self-similarity between analogues may be only approximate, as in the case of statistical self-similarity (Mandelbrot, 1983; Peitgen et al, 1992), or it may be more exact. Most quantitative tests involving masses, radii, spin periods, pulsation periods, magnetic dipole moments, etc. show agreement to a factor of 2 or better (Oldershaw, 1989a). The following definitive (i.e., unique, testable, *non-adjustable* and prior) predictions of the SSCM are relevant to the present discussion.

**5.2. Predicted GDMO mass function**

(i) The SSCM predicted that the main mass peak of the galactic DM population would occur at $\approx 0.15$ $M_o$ (Oldershaw, 1987). This is within the 0.1 $M_o$ to 0.4 $M_o$ range derived from observations, and it is virtually identical to the best fit values of Zhao and de Zeeuw (1998), and Jetzer (1998). The predicted GDMOs in the main peak would constitute $\approx 39\%$ of all DM objects, by numbers, and $\approx 69\%$, by mass.

(ii) A second major mass peak was predicted (Oldershaw, 1987) to be centered on $\approx 7 \times 10^{-5}$ $M_o$, and is consistent with empirical hints (see Figure 1) of a planetary-mass peak in the DMMF. This planetary-mass GDMO population would constitute $\approx 56\%$ of the total *number* of GDMOs, but $\ll 1\%$ of the DM mass. Parenthetically, the SSCM also anticipated the existence of pulsar/planet systems (Oldershaw, 1996).

(iii) The SSCM predicted (Oldershaw, 1989a,b) that there will be a low abundance region in the DMMF between $\approx 2 \times 10^{-4}$ $M_o$ and $\approx 0.1$ $M_o$, which may correspond to the observed "gap" in stellar scale objects between roughly $5 \times 10^{-3}$ $M_o$ and $0.1$ $M_o$.

(iv) Another small GDMO mass peak centered on $\approx 0.58$ $M_o$ was predicted (Oldershaw, 1989b) to constitute $\approx 4\%$ of all GDMOs, accounting for $\approx 30\%$ of the total GDMO mass. The remaining 1% of the galactic DM mass is predicted to be comprised of more massive GDMOs, with preferred mass peaks roughly at multiples of 0.15 $M_o$.

(v)  It is an intrinsic prediction of the SSCM that there is one universal mass function for sufficiently representative GDMO samples (Oldershaw, 1989a,b; 1996), which would explain the similarity between the halo and bulge DMMFs.

(vi)  Another inherent prediction of the SSCM is that the overall Galactic SMF and the overall galactic DMMF are nearly equivalent (Oldershaw, 1989a,b; 1996).

**5.3.  Predicted physical states of GDMOs**

The SSCM predicts that the stellar-mass and planetary-mass GDMO populations are comprised of ultracompact objects, i.e., black holes (Oldershaw, 1987), and that they constitute virtually all of the galactic dark matter.  Such a vast population of ultracompact objects might have important implications for research on gamma-ray bursts, as proposed by Cline (1996), X-ray backgrounds and cosmic ray origins.

**6. Conclusions**

The identity of the dark matter, making up 90% to 99% of the mass of the universe, is surely one of the most critical question facing astrophysicists today.  Can we claim even a rudimentary understanding of the cosmos, let alone speak of "precision cosmology", if we do not know its fundamental composition?  Fortunately we are poised to solve the dark matter problem in the forseeable future.  A considerable amount of evidence has already been gathered, and the next generation of microlensing experiments will further elucidate the mass function of GDMOs.

In this paper GDMO mass estimates derived from microlensing data and reported through May 1999 have been reviewed, along with trends in the data and relevant uncertainties.  The most reasonable, although still tentative, conclusions resulting from seven years of efforts to identify the GDMOs have also been also reviewed.  The inferred characteristics are difficult to understand within the context of the standard models of cosmology and stellar evolution.  On the other hand, a fractal model involving discrete self-similarity had previously made unique predictions that are consistent with the unusual observational findings. Several definitive tests can show us whether this match between predictions and observations is a critical piece of evidence for deciphering the dark matter enigma, or a statistically improbable coincidence.

If the SSCM is a useful step towards an improved cosmological paradigm, and the principle of discrete self-similarity is a fundamental property of nature, then the predictions discussed in sections 5.2 and 5.3 will be verified, with increasing accuracy, by future observations. If these predictions are not upheld, then the SSCM is wrong. Another seven years of microlensing experiments should be sufficient to verify or falsify these predictions.

**Acknowledgement:** The author would like to thank Dr. Phillipe Jetzer of the University of Zurich for helpful criticism.

**TABLE 1**
**GALACTIC DARK MATER MASS ESTIMATES**

| | Est. Mass ($M_o$) | Date (Mo/Yr) | Comments | References |
|---|---|---|---|---|
| 1 | $5.5 \times 10^{-5}$ | 12/91 | Component A, QSO 2237+ 0305 | Webster, et al., 1991 |
| 2 | 0.12 | 10/93 | First MACHO event (halo) | Alcock, et al., 1993 |
| 3 | $10^{-5}$ | 10/93 | QSO variability (Bimodal mass distribution: planetary + stellar?) | Refsdal and Stabell, 1993 |
| 4 | 0.2 | 10/93 | EROS #1 and #2 (halo) | Aubourg, et al., 1993 |
| 5 | $10^{-4}$ | 11/93 | $10^{-4}$ $M_o$ seems to give the best fit; large uncertainty; also see Hawkins, 1996 for comments | Schneider, 1993 |
| 6 | 0.144 | 3/94 | Method of moments analysis of MACHO #1 + EROS #1, #2 | Jetzer and Masso, 1994 |
| 7 | 0.08 | 4/94 | MACHO #1 + EROS #1, #2 | Evans and Jijina, 1994 |
| 8 | 0.08 | 9/94 | Method of moments analysis of MACHO #1-#3 + EROS #1, #2 | Jetzer, 1994 |
| 9 | $10^{-7}$ | 10/95 | Reanalysis of EROS short-term data suggests *possibility* of several planetary-mass events | Kerins, 1995 |
| 10 | $10^{-3}$ | 2/96 | QSO variability | Hawkins, 1996 |
| 11 | 0.08 | 4/96 | MACHO #1-#3 | Alcock, et al., 1996 |
| 12 | 0.27 | 5/96 | MACHO #1-#8 + EROS #1, #2 | Jetzer, 1996 |

| 13 | $10^{-5}$ | 6/96 | QSO variability, strong peak in planetary-mass range | Schild, 1996 |
|---|---|---|---|---|
| 14 | 0.50 | 6/96 | MACHO #1-#8 | Pratt, et al., 1996 |
| 15 | 0.40 | 8/96 | MACHO #1-#7, EROS #1, #2 | Flynn et al., 1996 |
| 16 | $10^{-5.5}$ | 9/96 | Second analysis of No. 5 above | Schild and Thompson, 1996 |
| 17 | $\sim 10^{-3}$ | 6/97 | QSO microlensing analysis, <m> | Hawkins and Taylor, 1997 |
| 18 | 0.27 | 6/97 | Halo events, $<m_h>$ | De Paolis, et al., 1997 |
| 19 | 0.26 | 6/97 | 16 halo events | Jetzer, 1997 |
| 20 | 0.16 | 6/97 | ~150 bulge events | Jetzer, 1997 |
| 21 | 0.5 (0.04-0.8) | 9/97 | MACHO 2-yr report, "most probable" mass, MACHO #1-#8 + EROS #1, #2 | Alcock, et al., 1997b |
| 22 | 0.05-1.0 | 2/98 | Uncertainty of ~ 4, due to model uncertainties | De Paolis, et al., 1998 |
| 23 | ~ 0.15 | 2/98 | MACHO bulge events, 1st yr, ~40 events | Mera, et al., 1998 |
| 24 | 0.4 – 0.5 | 2/98 | MACHO halo events, 2-yr, ~10 events | Mera, et al., 1998 |
| 25 | 0.15 – 0.4 | 3/98 | $<m_h>$, testing model dependencies | Gyuk and Gates, 1998 |
| 26 | 0.3 (0.2-0.36) | 6/98 | $<m_h>$, thick disk + spheroid model | Gates, et al., 1998 |

| 27 | 0.15 | 6/98 | ~ 200 MACHO bulge events | Zhao and de Zeeuw, 1998 |
| --- | --- | --- | --- | --- |
| 28 | $2.5 \times 10^{-6}$ to $1.4 \times 10^{-2}$ | 8/98 | QSO 0957+561A,B | Pelt, et al., 1998 |
| 29 | $6 \times 10^{-68}$ to $5 \times 10^{-65}$ | 8/98 | Neutrino oscillation report | Fukaday, et al., 1998 |
| 30 | 0.3 | 8/98 | General halo average | Evans and Gyuk, 1998 |
| 31 | ~ 0.40 | 9/98 | General MACHO + EROS <m> | Alfonso, et al., 1998 |
| 32 | 0.23 (0.2 – 0.3) | 10/98 | 2 SMC + 1 LMC events | Sahu, 1998 |
| 33 | 0.1 - 1.0 | 10/98 | OGLE data estimates | Han and Chang, 1998 |
| 34 | 0.1 - 0.6 | 10/98 | MACHO results, 2-yr | Markovic, 1998 |
| 35 | 0.5 (0.3-0.8) | 12/98 | MACHO group update | Sutherland, 1999 |
| 36 | ~ 0.1 | 12/98 | Review of bulge and halo data | Jetzer, 1998 |
| 37 | ~ $5 \times 10^{-56}$ | 1/99 | Possible evidence for WIMP | Glanz, 1999 |
| 38 | 0.26 (to 0.50) | 1/99 | Halo events to date | Jetzer, 1999 |
| 39 | ~ $10^{-3}$ | 1/99 | QSO microlensing | Hawkins, 1999 |
| 40 | 0.3 | 3/99 | Microlensing review | Sadoulet, 1999 |

| 41 | 0.3 | 3/99 | EROS II, 3 spiral arm events | Derue, et al., 1999 |
| 42 | ~ $10^{-5}$ to $10^{-4}$ | 5/99 | 1$^{st}$ planetary-mass MACHO event | Rhie, et al., 1999 |

**FIGURE CAPTIONS**

**Table 1. Galactic Dark Matter Mass Estimates**

Sequentially numbered mass estimates for galactic dark matter objects are given in units of $M_o$ (solar masses), along with their month/year of publication. Relevant comments and references are listed for each estimate.

**Figure 1. Dark Matter Mass Data vs Publication Date**

The full mass range for potential galactic dark matter objects (GDMOs) is shown along the x-axis in solar masses. The y-axis represents the number of months since November 1991, when positive GDMO mass estimates were first reported. The two vertical lines on the right side of the graph are the primary GDMO mass peaks predicted by the Self-Similar Cosmological Model, a discrete fractal model of the cosmos. Columns on the left side of the graph are the most likely DM mass ranges suggested by the Big Bang + Inflation Model. The horizontal dotted line at 57 months is the dividing line between new and previously reviewed mass estimates. Reported GDMO mass data from Table 1 appear to form separate clusters in the planetary-mass and stellar-mass ranges. These clusters may be correlated with the SSCM predictions.

**Figure 2. MACHO-Range DM Mass Data**

An expanded view of the right-hand portion of Figure 1.

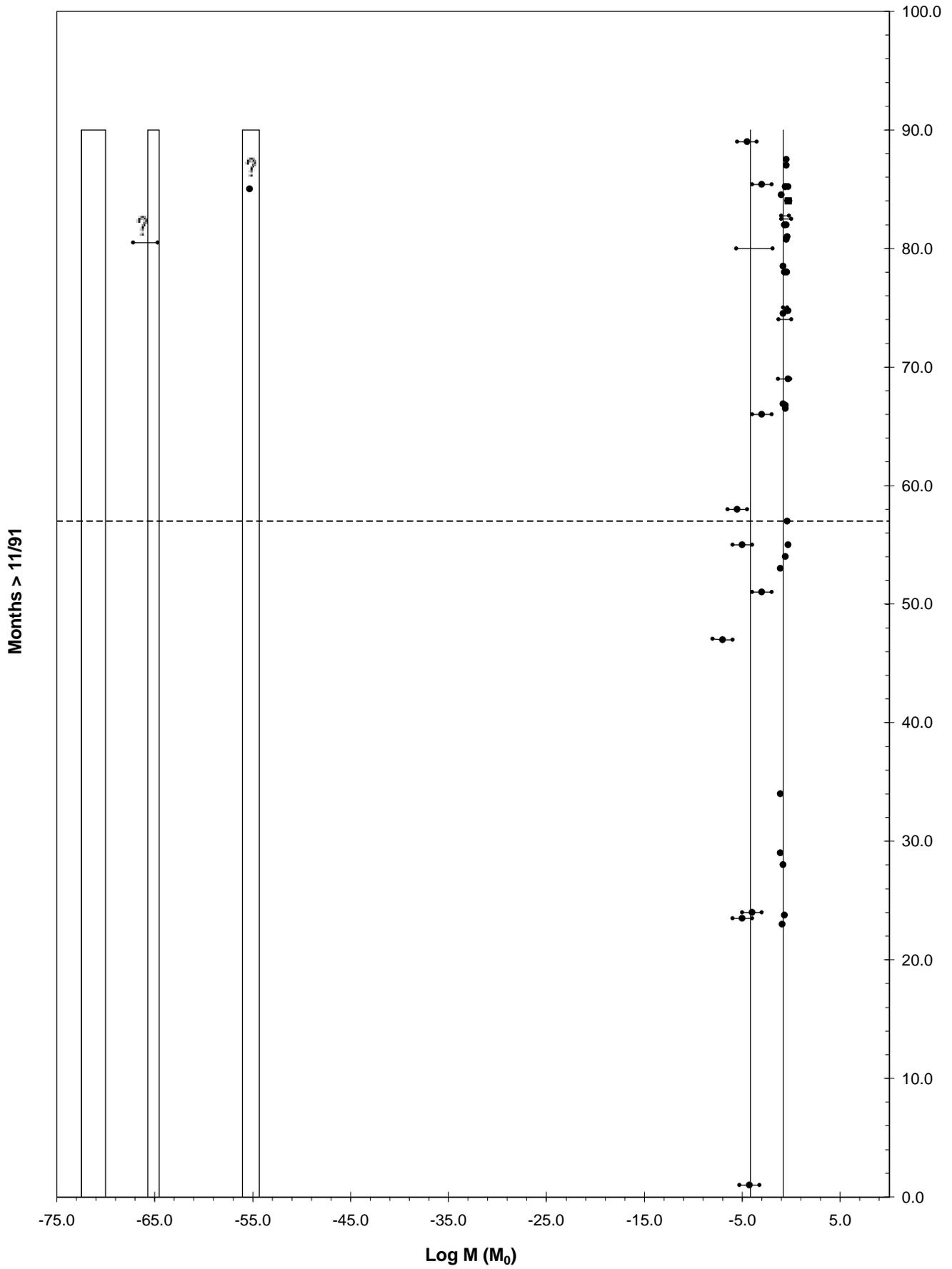

**Figure 1**

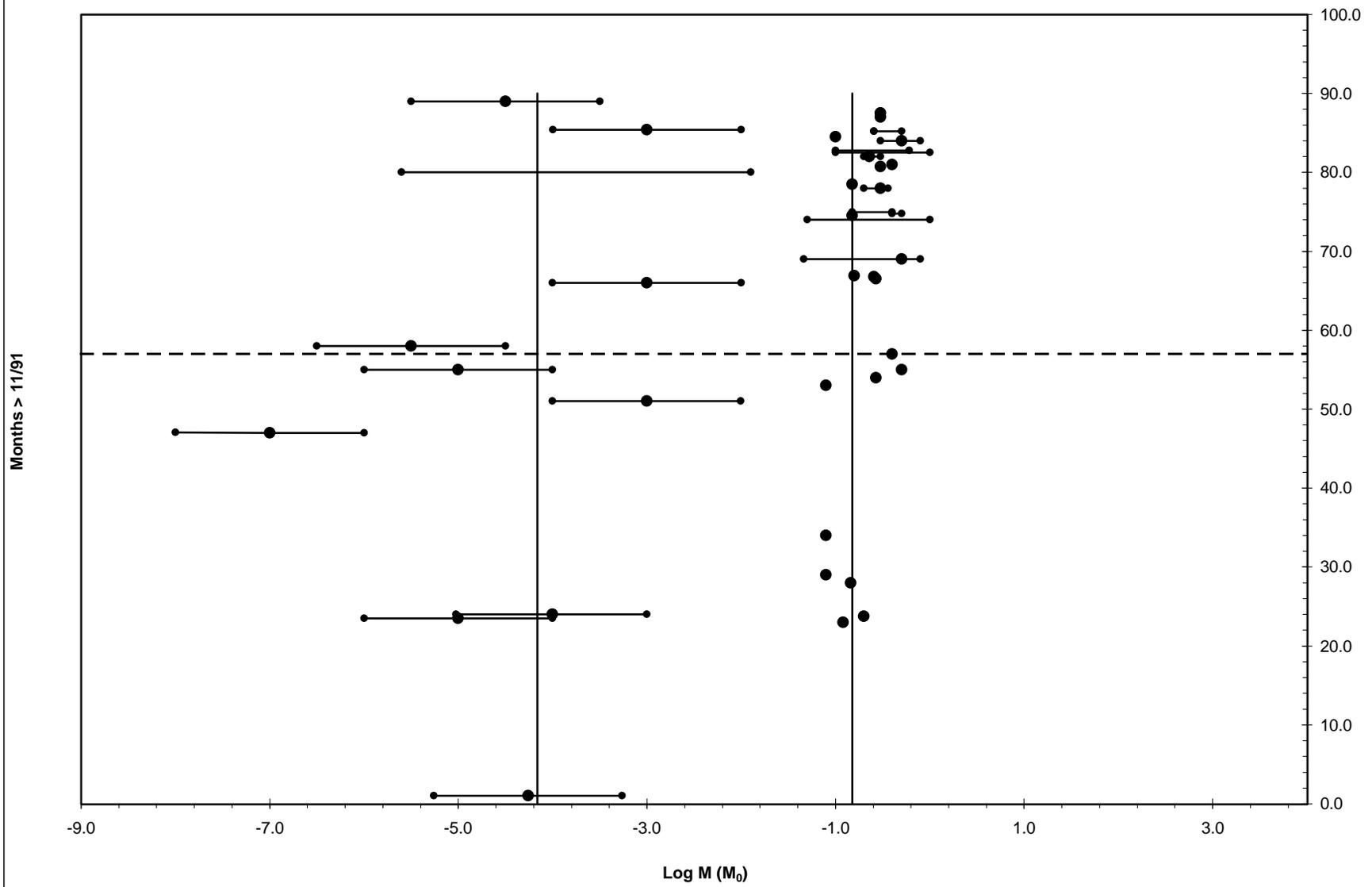

Figure 2